\documentclass[aps,prd,amsfonts,amssymb,preprint,eqsecnum,nofootinbib,superscriptaddress]{revtex4}
\usepackage{amsmath, mathrsfs}
\usepackage{hyperref, graphicx, subfigure}
\usepackage{ulem}

\newcommand\strk[1]{\textrm{\it{\sout{d}}}}
\newcommand\dslash{\hspace{-6pt}/}
\begin{document}

\title{
Weighted Power Counting and Perturbative Unitarity
\vskip 0.1in}
\author{Dylan Albrecht}

\affiliation{Particle Theory Group, Department of Physics,
College of William and Mary, Williamsburg, VA 23187-8795}
%

\newcommand\sect[1]{\emph{#1}---}

\begin{abstract}
We consider the relationship between renormalizability and unitarity at a
Lifshitz point in \(d\) dimensions.
We test tree unitarity for theories containing only scalars and
fermions, and for pure gauge theory.  In both cases, we find the
requirement of weighted power-counting renormalizability is equivalent to that
of tree unitarity.
\end{abstract}
\maketitle

\section{Introduction}
Lorentz violating (LV) field theories have been studied 
extensively, with constraints being placed on Lorentz violating
operators of the Standard Model (see, for example, 
Refs. \cite{Colladay:1998fq,Kostelecky:2010ux}). 
The idea of breaking Lorentz invariance by imposing
Lifshitz-point scaling opened the doors to rendering previously
nonrenormalizable theories renormalizable \cite{Anselmi:2007ri}.
The main incentive is to obtain an UV (albeit Lorentz violating) completion
of a nonrenormalizable field theory which becomes Lorentz invariant in the
infrared.  There are some advantages to invoking Lorentz violation. 
From the point of view of
eliminating unwanted ultraviolet divergences, there are many regularization
techniques available.  In each technique, the regularization is usually removed
in some manner, but Lorentz violation provides a 
physical cutoff \cite{Visser:2009fg}.  Also, by imposing
Lifshitz-point scaling, one can make virtually any theory power-counting
renormalizable.  Unfortunately, this is not a panacea as, for instance, there
is no \textit{a priori} equivalence between power-counting renormalizability and
unitarity. For example, the Standard Model is power-counting renormalizable,
but one can derive perturbative unitarity bounds on the Higgs mass.
The preceding remarks are from the point of view of a Wilsonian QFT.
It is interesting to note, as recently proposed by Dvali \textit{et al.} 
\cite{Dvali:2010jz,Dvali:2010ns}, it may be possible to have nonrenormalizable,
strongly coupled theories which self-unitarize by formation of extended,
classical field configurations.  Thus, the indication of strong coupling
does not necessarily imply new physics, but the theory may begin to obstruct
short distance measurements in analogy to the formation of black holes
in two-to-two scattering at trans-Planckian energy.

Most recently, Lifshitz-point field theories
have gained popularity because of the prospect of producing a consistent,
renormalizable quantum theory of gravity \cite{Anselmi:2007ri, Horava:2009uw}.
A Lifshitz point is a conformal fixed point invariant under anisotropic
rescalings of space and time, with suitable scaling dimensions for fields.
The anisotropic scaling leads to a modification of power counting arguments
for renormalizability, and also changes the relativistic phase space
factor thereby altering the condition for perturbative unitarity.
Many theories can be constructed in which Lifshitz-point scaling restores
renormalizability; for example, consider gauge theories in higher dimensions
\cite{Anselmi:2008bq,Anselmi:2008bs,Anselmi:2009ng,Chen:2009ka,Iengo:2010xg}.
Unfortunately, Lifshitz type UV-completions of these theories are not necessarily without problems.  The proposed UV completion of 5d QED exhibits a fine
tuning problem
\cite{Iengo:2010xg}, and in the case of Ho\v{r}ava-Lifshitz gravity there has
been concern over the consistency of various versions of the theory
\cite{Blas:2009yd,Charmousis:2009tc,Blas:2009qj,Koyama:2009hc,Papazoglou:2009fj,
Blas:2009ck}.
Some of these versions become strongly coupled at a certain scale and there
is a breakdown of the perturbative expansion (for recent reviews related
to this problem see \cite{Sotiriou:2010wn,Padilla:2010ge}).
One way to see this breakdown is to check the bound for perturbative unitarity
\cite{Blas:2009yd, Cornwall:1974km}.  Since making a theory power-counting
renormalizable does not guarantee the absence of strong coupling, it
is interesting to ask what happens to perturbative unitarity for an arbitrary
theory at a Lifshitz point.

The purpose of this paper is to present, in a simple setting, the
manner in which making a theory renormalizable affects perturbative unitarity--
in particular, perturbative unitarity at tree level 
\cite{Cornwall:1974km}. We will quickly review some necessary background
material for a theory containing scalars and fermions, and then for a
pure gauge theory.  Then, we will derive the condition for tree unitarity
in tree-level scattering processes, and apply it in these two settings.

\section{Background}
We attempt to succinctly present the relevant material on scalars, fermions,
and gauge fields at a Lifshitz point.  A more complete
story of scalars and fermions can be found in Ref. \cite{Anselmi:2007ri}, and for
a more detailed discussion of gauge fields, see Refs.
\cite{Anselmi:2008bq, Anselmi:2008bs}.  We will, for the most part, follow
the notation of \cite{Anselmi:2008bs},
where we consider a spacetime manifold of dimension \(d\) to be split
as the product \(\mathbb{R} \times \mathcal{M}_{\bar{d}}\).  The spatial
manifold \(\mathcal{M}_{\bar{d}}\) is of dimension \(\bar{d}\) and the symmetry
group considered as \(O(\bar{d})\).  In general, we can consider the
spacetime manifold to be split into two sets of coordinates.
If we assume time and some spatial coordinates
to be in the first set, then the second set contains only spatial coordinates.
When appropriate, we will use a hat to denote the set of coordinates
containing time and a bar to denote the remaining spatial coordinates.
So, for instance, the dimension
of the spacetime is \(d = \hat{d} + \bar{d}\).  As evidenced above, we will
work in the special case of \(\hat{d} = 1\), where time is split from the
spatial coordinates.  The case
\(\hat{d} = 1\) is important because it is contained in a set of sufficient
conditions for
the absence of spurious subdivergences, as described in section III.
Also, \(\hat{d} = 1\) is the case considered for Ho\v{r}ava-Lifshitz gravity.

Scaling at a Lifshitz point by the parameter \(\lambda\) results in the
transformation
\begin{equation}
\hat{x} \rightarrow \lambda \hat{x}, \qquad \bar{x} \rightarrow \lambda^{1/z}
  \bar{x},
\end{equation}
where \(z\) is a positive nonzero integer representing the severity of the
difference in scaling.  For this to be a symmetry of the action the
fields must scale accordingly.

\subsection{Scalars and Fermions}
For the purposes of this paper, the free part of the Lagrangian for a scalar
and fermion can be written as:
\begin{equation}
\mathcal{L}_{free} = \frac{1}{2} (\hat{\partial} \phi)^2
  + \frac{1}{2 \Lambda_{L}^{2 z - 2}} \left(\bar{\partial}^{z} \phi \right)^2
  + \bar{\psi} i\hat{\partial}\dslash \psi
  + \frac{1}{\Lambda_{L}^{z - 1}} \bar{\psi}
    (i \bar{\partial \dslash})^{z} \psi.
\end{equation}
We have made use of some short-hand notation, which can be written out explicitly
as
\begin{equation*}
(\hat{\partial} \phi)^2 = \sum_{i,j}^{\hat{d}} (\hat{\partial}_{i} \phi)
    (\hat{\partial}_{j} \phi) \eta^{ij},
  \quad \textrm{and} \quad \left(\bar{\partial}^{z} \phi \right)^2 =
    \sum_{\substack{(i_{1}, \cdots, i_{z}) \\ (j_{1},
      \cdots, j_{z})}}^{d} (\bar{\partial}_{i_{1}} \cdots
    \bar{\partial}_{i_{z}} \phi) (\bar{\partial}_{j_{1}} \cdots
    \bar{\partial}_{j_{z}} \phi) \eta^{i_{1} j_{1}} \cdots \eta^{i_{z} j_{z}},
\end{equation*}
where the indices of the first sum start from one and the indices of the
second sum all start from
\(\hat{d} + 1\).  The tensor \(\eta\) is the \(d\)-dimensional Minkowski
metric with components \(\eta_{11} = 1\), \(\eta_{ii} = -1\) for \(i > 1\),
and the rest are zero.  Short-hand notation was also used to write the fermion
part of the Lagrangian, with contractions between partial derivatives and
gamma matrices, but we omit the explicit form as it is clear from the above
scalar example. Finally, the parameter \(\Lambda_{L}\) dictates the energy at
which the anisotropic scaling is important.  If we assign the weighted dimensions
\begin{equation}
[\hat{\partial}] = 1, \qquad [\bar{\partial}] = \frac{1}{z},
\end{equation}
we see that the weighted dimension of the spacetime volume element
\([d^d x] = [d\hat{x} d^{\bar{d}} \bar{x}] = - 1 - \bar{d}/z \equiv -\,\strk{d}\).
Thus, the weighted dimension of the Lagrangian is \(\strk{d}\).  By comparison,
we also find the following assignments:
\begin{equation}
[\phi] = \frac{1}{2}(\strk{d} - 2), \qquad [\psi] = \frac{1}{2} (\strk{d} - 1).
\end{equation}
The propagator for the scalar field will take the following form:
\begin{equation}
\label{sprop}
i \Delta_{F}(p) = \frac{i}{\hat{p}^{2} -
    \bar{p}^{2z}/\Lambda_{L}^{2z-2}},
\end{equation}
and we see, as \([1/\hat{p}^{2}] = -2\), the weighted dimension (or weight)
of the propagator is minus two. Analogously, the weight of the fermion
propagator is minus one.

\subsection{Gauge Fields}
If we decompose the gauge field as \(A = (\hat{A}, \bar{A})\) and the covariant
derivative as \(D = (\hat{D}, \bar{D}) = (\hat{\partial} - ig\hat{A},
\bar{\partial} - ig\bar{A})\),
where \(g\) is the gauge coupling, we have the following weighted dimensions:
\begin{equation}
[g \hat{A}] = [\hat{D}] = 1, \qquad [g \bar{A}] = [\bar{D}] = \frac{1}{z}.
\end{equation}
We can also separate the field strength by its components, and make the
following short-hand definitions:
\begin{equation}
\hat{F} \equiv F_{\hat{\mu} \hat{\nu}}, \quad \tilde{F} \equiv
    F_{\hat{\mu} \bar{\nu}},    \quad \bar{F} \equiv F_{\bar{\mu} \bar{\nu}}.
\end{equation}
For the case where \(\hat{d} = 1\), we have that \(\hat{F}\) is identically
zero, but we will temporarily assume the case of general
\(\hat{d}\) to determine the weight assignments.
If we consider the term \((\hat{\partial}\hat{A})^{2}\) to be of weight
\(\strk{d}\), then we can determine the weight of the gauge coupling
\([g] = 2 - \strk{d}/2\), and the weights of the gauge fields and field
strength components:
\begin{equation}
[\hat{A}] = \frac{\strk{d}}{2} - 1, \quad [\bar{A}] = \frac{\strk{d}}{2} - 2
  + \frac{1}{z}, \quad [\hat{F}] = \frac{\strk{d}}{2}, \quad [\tilde{F}] =
    \frac{\strk{d}}{2} - 1 + \frac{1}{z}, \quad [\bar{F}] = \frac{\strk{d}}{2}
  - 2 + \frac{2}{z}.
\end{equation}
Also, for later calculations, the weights of the propagators are
\cite{Anselmi:2008bq}:
\begin{equation}
\hat{P} \equiv [\langle \hat{A} \hat{A} \rangle] = -2,
    \quad \tilde{P} \equiv [\langle \hat{A} \bar{A} \rangle] = -3 + \frac{1}{z},
    \quad \bar{P} \equiv [\langle \bar{A} \bar{A} \rangle] = -4 + \frac{2}{z}.
\end{equation}

We will only be concerned with cases where the couplings appearing in
interactions,
\(\lambda_{i}\), have positive weight.  In particular, we wish to investigate
the class of theories which have all \([\lambda_{i}] \ge \chi\), where \(\chi\)
is some non-negative, minimal weight and the Lagrangian is written as
\begin{equation}
\label{glag}
\mathcal{L} = \frac{1}{\bar{g}^{2}}
    \mathcal{L}_{r}(\bar{g}A, \bar{g}\bar{C}, \bar{g}C),
\end{equation}
where \(C\) and \(\bar{C}\) denote the ghosts and antighosts.
The coupling \(\bar{g}\) (not necessarily the gauge coupling) is a factor of
the interaction couplings,
\(\lambda_{i} = \bar{\lambda}_{i} \bar{g}^{n_{i} - 2}\), and has weight
\begin{equation}
[\bar{g}] = \min_{i} \frac{[\lambda_{i}]}{n_{i} - 2},
\end{equation}
where \(n_{i}\) corresponds to the \(i\)-th vertex with \(n\) external legs
(\(n > 2\)).  The weight of \(\bar{g}\) satisfies the relations
\begin{equation}
[\lambda_{i}] \ge (n_{i} - 2) [\bar{g}],
    \quad \textrm{and} \quad 0 \le [\bar{g}] \le [g],
\end{equation}
such that \([\bar{\lambda}_{i}] \ge 0\).  Weighted power-counting
renormalizable Lagrangians of the form in \eqref{glag} have been proven to
be renormalizable (see Ref. \cite{Anselmi:2008bs}).

We note that the hat component of the gauge field has the
same weight as the scalar field, while the bar component has lower weighted
dimension. In some instances, the weight of \(\bar{F}\) can even be negative.
If we write the vertices as products of \(\bar{g} F\) and covariant derivatives,
\(\bar{F}\) may have negative weight while 
preserving polynomiality of the Lagrangian.  In order to have a finite number
of interaction terms, we require \([\bar{g} \bar{F}] > 0\), as this covers the
other components of \(F\) as well.  Thus, \([\bar{g}]\) is bounded above and
below:
\begin{equation}
-[\bar{F}] < [\bar{g}] \le [g].
\end{equation}
Of course, if \([\bar{F}]\) is positive the lower bound is zero.
The range of possible values for the weight of \(\bar{g}\) will dictate the
set of allowed interactions; consequently, \([\bar{g}] = [g]\) is the most
restrictive.

\subsection{Power Counting}
We will now quickly review the method of weighted power counting for a single
field, as in Ref. \cite{Anselmi:2007ri}.  Consider a
diagram with \(E\) external legs, \(I\) internal lines, \(L\) loops, and \(V\)
vertices.  In general, the
diagram will involve an integral of the form:
\begin{equation}
\prod_{i}^{L}\left( \int d \hat{q_{i}} d^{\bar{d}} \bar{q_{i}} \right)
\prod_{j}^{I} P_{j} \prod_{k}^{V} V_{k},
\end{equation}
where \(P_{i}\) are the propagators on the internal lines and \(V_{k}\) are
the vertices in the diagram.  If a vertex contains \(n\) hat derivatives
and \(m\) bar derivatives, we define the weighted degree of divergence of an
\(N\)-point vertex of type \(\alpha\) as \(\delta_{N}^{(\alpha)} = n + m/z\).
We also define the number of vertices, \(v_{N}^{(\alpha)}\),
corresponding to an \(N\)-point interaction of type \(\alpha\).  The weighted
superficial degree of divergence (\(\omega\)) can be written as the sum of
the contribution from the loop measure, propagators, and \((\alpha, N)\)-type
vertices carrying momentum factors of weight \(\delta_{N}^{(\alpha)}\).
\begin{equation}
\omega = L\strk{d} + PI + \sum_{(\alpha, N)} \delta_{N}^{(\alpha)}
    v_{N}^{(\alpha)},
\end{equation}
where the weight of the propagator is \(P\) and the final term is the sum over
all the vertices in the diagram.  Using the topological relations
\(L = I - V + 1\) and \(E + 2I = \sum N v_{N}^{(\alpha)}\), we arrive at the
expression
\begin{equation}
\omega = L\strk{d} -\frac{E}{2}(\strk{d} + P)
  + \sum v_{N}^{(\alpha)} (\delta_{N}^{(\alpha)} - D(N)),
\end{equation}
where
\(D(N) \equiv \strk{d}(1 - \frac{N}{2}) - P\frac{N}{2} = \strk{d}
- \frac{N}{2}(\strk{d} + P)\).
Now, the condition for weighted
power-counting renormalizability is \(\delta_{N}^{(\alpha)} \le D(N)\).  This
relation implies there are no couplings of negative weighted dimension.
Likewise, it implies there are no operators of weighted dimension greater than
\(\strk{d}\).
Since we will deal in some detail with \(D(N)\), for two types of theories,
we show its resulting expression in each case.  Note,
for all fields (\(f\)) the dimensions of the fields may be written as
\([f] = \frac{1}{2}(\strk{d} + P_{f})\), where \(P_{f}\) is the weight of the
propagator of \(f\).  For theories only containing scalars and fermions, the
result for \(D(N)\) may be written as
\begin{equation}
D(N_{B} + N_{F}) = \strk{d} - N_{B} [\phi] - N_{F} [\psi],
\end{equation}
where \(N_{B}\) is the number of bosons and \(N_{F}\) is the number of fermions.
A similar expression is obtained for pure gauge theories:
\begin{equation}
D(\hat{N} + \bar{N} + N_{gh}) = \strk{d} - \hat{N} [\hat{A}] - \bar{N} [\bar{A}]
  - N_{gh} [C],
\end{equation}
where \(\hat{N}\) is the number of hat-component gauge fields, \(\bar{N}\)
is the number of bar-component gauge fields, and \(N_{gh}\) is the number of
ghosts and antighosts.  Since we will only be concerned with tree-level diagrams,
we set \(N_{gh} = 0\).

\section{Perturbative Unitarity Condition}
In order to determine the condition for perturbative unitarity, we proceed
by developing the formalism in analogy to the more familiar discussion in
four dimensions maintaining Lorentz invariance (a similar derivation, scattering
scalars in 4d at a Lifshitz point, was found in \cite{Blas:2009ck}).
We may start with the expression of the generalized optical theorem
for forward scattering \cite{Peskin:1995ev}:
\begin{equation}
\label{opthm}
2 Im\left[\mathcal{M}(k_{1} k_{2} \rightarrow k_{1} k_{2} )\right] =
  \sum_{n} \int d\Pi_{n} \left|\mathcal{M}(k_{1} k_{2} \rightarrow
  \{q_{n}\})\right|^{2}.
\end{equation}
Labelling the initial state as '\(a\)' and separating out the elastic portion,
we have
\begin{equation}
  \label{ucond}
2 Im\left[\mathcal{M}(a \rightarrow a)\right] - \int
  \frac{d^{\bar{d}} \bar{q}_{1} d^{\bar{d}} \bar{q}_{2}}
    {(2 \pi)^{2 \bar{d}} E_{1} E_{2}}
  \left|\mathcal{M}(a \rightarrow q_{1} q_{2})\right|^2 (2\pi)^d
  \delta^{(d)}(k_{1} + k_{2} - q_{1} -q_{2}) > 0.
\end{equation}
To proceed with the derivation, we presume the scattering takes place in the
center-of-mass frame.  Assuming we have a dispersion relation that looks
like \(E = \sqrt{f(\bar{q}) + m^2}\), where \(f(\bar{q})\)
is a positive, monotonic function of the magnitude of the spatial momenta,
we can perform most of the  integrals to get:
\begin{equation}
\frac{\bar{q}_{1}^{\bar{d} - 1}}{(2 \pi)^{\bar{d} - 1} 4 E_{cm} f'(\bar{q}_{1})}
  \int d\Omega_{\bar{d} - 1} \left|\mathcal{M}\right|^{2} \sim
  \frac{1}{4 (2 \pi)^{\bar{d} - 1}}
  E_{cm}^{\bar{d}/z - 3} \int d\Omega_{\bar{d} - 1}\left|\mathcal{M}\right|^{2},
\end{equation}
where we have taken \(f'(\bar{q}) \sim \bar{q}^{2z - 1}\), at high energy.
In general, for two-to-two scattering
\(\lambda_{1} \lambda_{2} \rightarrow \lambda_{3} \lambda_{4}\), \(\mathcal{M}\)
is the helicity amplitude
\(\mathcal{M}_{\lambda_{1} \lambda_{2}, \lambda_{3} \lambda_{4}}\),
where \(\lambda_{i}\) corresponds to the helicity of the \(i\)-th particle.
The scattering takes place in a plane, and
the amplitude is a function of \(E_{cm}\) and
the angle \(\theta\) between incoming and outgoing particles.
The helicity amplitude can then be expanded in terms of Wigner d-functions:
\(d_{\Lambda \Lambda^{'}}^{j}(\theta)\), with
\(\Lambda = \lambda_{1} - \lambda_{2}\) and
\(\Lambda^{'} = \lambda_{3} - \lambda_{4}\).  In the following,
we assume specific helicity configurations such that
\(\Lambda = \Lambda^{'} = 0\),
where the d-functions become the Legendre polynomials:
\(d_{00}^{j}(\theta) = P_{j}(\cos(\theta))\).
This is done for clarity of presentation, but it should be possible to
generalize the result to arbitrary helicity considerations.
Now, we expand the invariant scattering amplitude in terms of Legendre
polynomials:
\begin{equation}
\mathcal{M}(E_{cm}, \cos(\theta)) = 16 \pi \sum_{j} (2j + 1) a_{j}^{el}
  P_{j}(\cos(\theta)).
\end{equation}
Plugging this into \eqref{ucond}, we get the following expression:
\begin{equation}
\label{jsumineq}
32 \pi \sum_{j} (2j + 1) Im(a_{j}^{el}) P_{j}(\cos(\theta))
  - C(\bar{d}) E_{cm}^{\bar{d}/z - 3} \sum_{j} (2j + 1) \left|a_{j}^{el}\right|^2
  > 0,
\end{equation}
where \(C(\bar{d})\) is a constant, which depends on \(\bar{d}\),
resulting from the various integrations.
Since the scattering matrix for elastic scattering is diagonal in \(j\),
equation \eqref{jsumineq} constrains each partial-wave amplitude \(a_{j}^{el}\)
independently.  After some rearranging, we arrive at the following:
\begin{equation}
Re(a_{j}^{el})^2 + \left[Im(a_{j}^{el})
  - \frac{16 \pi}{C(\bar{d})} E_{cm}^{-(\bar{d}/z - 3)}\right]^2 <
  \left[\frac{16 \pi}{C(\bar{d})} E_{cm}^{-(\bar{d}/z -3)}\right]^2.
\end{equation}
The above inequality defines the unitarity circle; as long as we are within
the circle, perturbative unitarity holds.  This translates into a bound on
the energy growth of the right-hand side of equation \eqref{opthm}:
\begin{equation}
  \label{ebound}
  \frac{1}{2} \sum_{n} \int d\Pi_{n} \left|\mathcal{M}(a \rightarrow
  \{q_{n}\})\right|^{2} \; = \; Im\left[\mathcal{M}(a
  \rightarrow a)\right] \; \lesssim \; (\textrm{const})\, E^{-(\strk{d}\, - 4)}.
\end{equation}
The condition for tree unitarity then follows from some dimensional
analysis. Using
\begin{equation}
\left[d\Pi_{n}\right] = n\left(\strk{d} - 2\right) - \strk{d},
  \quad \textrm{and assuming} \quad \mathcal{M} \sim E^{\beta},
\end{equation}
we get, from the energy bound \eqref{ebound},
\begin{equation}
\label{betaucond}
\beta \le 2 - \frac{n}{2} \left(\strk{d} - 2\right) = \strk{d}
     - \frac{N}{2}(\strk{d} - 2),
\end{equation}
where, in the final equality, we substituted \(n = N-2\).

\subsubsection{Application to Scalars and Fermions}
In order to check the condition of tree unitarity for scalars and fermions,
it is useful to rewrite the unitarity condition as
\begin{equation}
\label{rewriteucond}
\beta \le D(N) + \sum_{i}^{N} \delta(f_{i}),
\end{equation}
where the sum is over external lines, and \(\delta(f_{i})\) is the
highest power of energy the field \(f_{i}\), when contracted with an external
state, can contribute to the scattering amplitude.  A scalar external line
contributes an energy of \(E^{0}\), while a fermion external line contributes,
at most, \(E^{1/2}\).  For example, the four-point interaction
\(\mathcal{L} \supset \phi \phi \bar{\psi} \psi\) has
\(\sum \delta(f) = 2 \delta(\phi) + 2 \delta(\psi) = 2 (0) + 2 (1/2) = 1\).
Thus, the tree-level scattering amplitude grows at most like \(E\).
Now, consider the general interaction term written schematically as
\begin{equation}
\label{schmint}
k \hat{\partial}^{s} \bar{\partial}^{t} \phi^{N_{B}} (\bar{\psi} \psi)^{N_{F}/2},
\end{equation}
where \(N_{B}\) counts the number of scalars, \(N_{F}\) is the number of
fermions, \(t+s\) is the number of derivatives, and \(k\) is a constant of
dimensionality \(\kappa\). We should note, perturbative unitarity can also be
violated if the propagator contains more than two time derivatives. For the
argument that there are no more than two time derivatives, and in particular
no time derivatives in interactions with \(N > 2\), the reader may check Ref.
\cite{Anselmi:2007ri}. The weighted degree of divergence of the
(\(N_{B} + N_{F}\))-point interaction in equation \eqref{schmint} is
\(\delta_{N_{B} + N_{F}} = s + t/z\). The contribution
from the external lines can be represented as \(\sum \delta(f)\), as
defined before.  Substituting \(\beta = \delta_{N_{B} + N_{F}} + \sum \delta(f)\)
into equation \eqref{rewriteucond} we get
\begin{equation}
\label{npoint}
\delta_{N_{B} + N_{F}} \le D(N_{B} + N_{F}),
\end{equation}
which is the condition for weighted renormalizability from before.  So, for
an \(N\)-point interaction, the unitarity condition is equivalent to
the renormalizability condition.

To check tree unitarity for a tree-level diagram containing a propagator,
equation \eqref{rewriteucond} is again the most convenient.
This condition, for a
vertex with \(N_{1}\) lines connected to a vertex with \(N_{2}\) lines
by the field \(f_{prop}\) with a propagator of weight \(P\), is
\begin{equation}
\label{npropnbig}
\delta_{N_{1}} + \delta_{N_{2}} + P + \sum_{i}^{N_{1}} \delta(f_{i})
  + \sum_{i}^{N_{2}} \delta(f_{i}) - 2 \delta(f_{prop}) \le D(N_{1} + N_{2} - 2)
  + \sum_{i}^{N_{1} + N_{2} - 2} \delta(f_{i}),
\end{equation}
where \(\delta(f_{prop})\) is the energy factor the field \(f_{prop}\) would
contribute were it an external line.  We can expand
\(D(N_{1} + N_{2} - 2) = D(N_{1}) + D(N_{2}) + D(-2) - 2 \strk{d}\), and use
the fact that \(D(-2) = 2 \strk{d} + P\) to arrive at the condition:
\begin{equation}
\label{npropn}
\delta_{N_{1}} + \delta_{N_{2}} \le D(N_{1}) + D(N_{2}),
\end{equation}
which always holds since the individual vertices are renormalizable.
The result \eqref{npropn},
along with the result of \eqref{npoint}, implies that, for scalars and fermions
in tree-level scattering processes, the tree unitarity condition is equivalent
to the condition of weighted power-counting renormalizability.

\subsubsection{Application to Gauge Fields}
The treatment of gauge theory is arguably more interesting than that of
scalars and fermions.  For instance, in a 4d Lorentz invariant theory, we
cannot simply add a mass term for a gauge field, as the resulting theory would
violate unitarity. After witnessing the troubles present in the original
version of Ho\v{r}ava-Lifshitz gravity, it is natural to wonder, from the 
perspective of obtaining an UV-complete, higher dimensional gauge theory, what
happens to perturbativity.

For gauge fields at tree level, the analysis is analogous to the treatment for
scalars and fermions above, so we will briefly reiterate the arguments.
The condition of tree unitarity can again be
written in the form of equation \eqref{rewriteucond}, where the external line
contributions are \(\delta(\hat{A}) = 0\) and
\(\delta(\bar{A}) = -1 + \frac{1}{z}\).  We consider the following schematic
\(N\)-point vertex:
\begin{equation}
\bar{\lambda}_{i} \bar{g}^{N - 2} \hat{\partial}^{s} \bar{\partial}^{t}
    \hat{A}^{\hat{N}} \bar{A}^{\bar{N}},
\end{equation}
where \(\hat{N}\) is the number of \(\hat{A}\)s and \(\bar{N}\) is the number of
\(\bar{A}\)s.  We may write
\(\beta = \delta_{\hat{N} + \bar{N}} + \sum \delta(f)\), and from equation
\eqref{rewriteucond} we obtain
\begin{equation}
\label{gaugenpoint}
\delta_{\hat{N} + \bar{N}} \le D(\hat{N} + \bar{N}),
\end{equation}
which is the condition for power-counting renormalizability.
Similarly for \(N_{1}\)-point and \(N_{2}\)-point vertices connected by
a field with propagator of weight \(P\), we arrive at the following expression:
\begin{equation}
\delta_{N_{1}} + \delta_{N_{2}} + P \le D(N_{1} + N_{2} - 2).
\end{equation}
Since the propagator could be \(\langle \hat{A} \bar{A} \rangle\), we make
use of the relation \(\tilde{P} = \frac{1}{2} (\hat{P} + \bar{P})\). 
The result is the same as for the scalar and fermion case:
\begin{equation}
\label{gaugenpropn}
\delta_{N_{1}} + \delta_{N_{2}} \le D(N_{1}) + D(N_{2}),
\end{equation}
which holds if we assume each vertex is power-counting renormalizable.
So, at tree level, weighted power-counting renormalizable pure gauge theories
satisfy perturbative unitarity.

\section{Conclusions}
We have seen that while imposing Lifshitz-point scaling can render a
theory renormalizable, it also
modifies the relativistic phase space factor and thereby the condition for 
perturbative unitarity. For the theories considered,
the tree unitarity condition holds if and only if the Lagrangian is weighted
power-counting renormalizable.
\begin{acknowledgments}
I would like to thank Joshua Erlich for suggesting that I work on this project
and for his insightful comments and encouragement.  This research
was supported under NSF Grant PHY-0757481.
\end{acknowledgments}

\end{document}